\documentclass[aps,prl,onecolumn,superscriptaddress,floatfix,nobalancelastpage]{revtex4-1}

\usepackage{graphicx,bm}
\usepackage{setspace}
\usepackage{hyperref}
\begin{document}

\title{Reduced Hall carrier density in the overdoped strange metal regime of cuprate superconductors}

\author{Carsten Putzke}
\affiliation{H.H. Wills Physics Laboratory, University of Bristol, Tyndall Avenue, BS8 1TL, United Kingdom}

\author{Siham Benhabib}
\affiliation{Laboratoire  National  des  Champs  Magn\'{e}tiques  Intenses  (LNCMI-EMFL),  (CNRS-INSA-UGA-UPS), Toulouse, 31400, France}

\author{Wojciech Tabis}
\affiliation{Laboratoire  National  des  Champs  Magn\'{e}tiques  Intenses  (LNCMI-EMFL),  (CNRS-INSA-UGA-UPS), Toulouse, 31400, France and
AGH University of Science and Technology, Faculty of Physics and Applied Computer Science, 30-059 Krakow, Poland.}

\author{Jake Ayres}
\affiliation{H.H. Wills Physics Laboratory, University of Bristol, Tyndall Avenue, BS8 1TL, United Kingdom}%

\author{Zhaosheng Wang}
\affiliation{Hochfeld-Magnetlabor Dresden (HLD-EMFL), Helmholtz-Zentrum Dresden-Rossendorf, D-01328 Dresden, Germany}

\author{Liam Malone}
\affiliation{H.H. Wills Physics Laboratory, University of Bristol, Tyndall Avenue, BS8 1TL, United Kingdom}%

\author{Salvatore Licciardello}
\affiliation{High Field Magnet Laboratory (HFML-EMFL) and Institute for Molecules and Materials,\\
Radboud University, Toernooiveld 7, 6525 ED Nijmegen, Netherlands}%

\author{Jianming Lu}
\affiliation{High Field Magnet Laboratory (HFML-EMFL) and Institute for Molecules and Materials,\\
Radboud University, Toernooiveld 7, 6525 ED Nijmegen, Netherlands}%

\author{Takeshi Kondo}
\affiliation{Institute for Solid State Physics, University of Tokyo, Kashiwa-no-ha, Kashiwa, Japan.}

\author{Tsunehiro Takeuchi}
\affiliation{Toyota Technological Institute, Nagoya 468-8511, Japan.}

\author{Nigel E. Hussey}%
\affiliation{H.H. Wills Physics Laboratory, University of Bristol, Tyndall Avenue, BS8 1TL, United Kingdom}
\affiliation{High Field Magnet Laboratory (HFML-EMFL) and Institute for Molecules and Materials,\\
Radboud University, Toernooiveld 7, 6525 ED Nijmegen, Netherlands}%

\author{John R. Cooper}
\affiliation{Department of Physics, University of Cambridge, Madingley Road, Cambridge, CB3 0HE, United Kingdom}

\author{Antony Carrington}%
\affiliation{H.H. Wills Physics Laboratory, University of Bristol, Tyndall Avenue, BS8 1TL, United Kingdom}%

\begin{abstract}

\end{abstract}

\maketitle

{\bf
Efforts to understand the microscopic origin of superconductivity in the cuprates are dependent on knowledge of the normal state. The Hall number in the low temperature, high field limit $n_{\rm H}(0)$ has a particular significance because within conventional transport theory it is simply related to the number of charge carriers, and so its  evolution with doping gives crucial information about the nature of the charge transport.  Here we report a study of the high field Hall coefficient of the single layer cuprates Tl$_2$Ba$_2$CuO$_{6+\delta}$ (Tl2201) and (Pb/La) doped Bi$_2$Sr$_2$CuO$_{6+\delta}$ (Bi2201) which shows how $n_{\rm H}(0)$ evolves in the overdoped, so-called strange metal, regime of cuprates.  We find that $n_{\rm H}(0)$ increases smoothly from $p$ to $1+p$, where $p$ is the number of holes doped into the parent insulating state, over a wide range of doping.  The evolution of $n_{\rm H}$  correlates with the emergence of the anomalous linear-in-$T$ term in the low-$T$ in-plane resistivity. The results could suggest that quasiparticle decoherence extends to dopings well beyond the pseudogap regime.}

In the search for the microscopic origin of high temperature superconductivity in the cuprates, much effort has been directed to understanding their normal state properties and how these are linked to superconductivity as a function of temperature and doping. The underdoped regime exhibits pseudogap phenomena as well as tendencies toward several types of order, or incipient order, including charge and spin density waves (CDW/SDW) \cite{Keimer15}. In the regime where CDW order has been detected, at low temperature and high fields, $n_{\rm H}$ changes sign \cite{Leboeuf07} suggesting some form of Fermi surface reconstruction \cite{Sebastian12}. In some overdoped cuprates, beyond the doping level $p^*$ where the pseudogap disappears,  there do not appear to be any competing orders and so potentially these materials provide a simpler starting point from which to understand the emergence of high temperature superconductivity.

In the far overdoped regime, the normal state behaviour of Tl2201 resembles, in many aspects, that of a conventional Fermi liquid with coherent quasiparticles around the entire Fermi-surface \cite{Vignolle08,Bangura10,Rourke10}, whose shape is found to be well described by conventional density functional theory \cite{Hussey03,Plate05}, albeit with a large (factor 3) renormalisation in the effective mass which derives from a narrowing of the band \cite{Bangura10,Rourke10}. One aspect of the overdoped regime which contrasts with that of conventional metals is the evolution of the in-plane resistivity $\rho_{xx}(T)$ with doping.   $\rho_{xx}(T)$ evolves smoothly from linear, close to optimal doping, to quadratic in the far overdoped regime \cite{Kubo91,Manako92} leading to this being called the `strange metal' regime \cite{Hussey08,Hussey13}.

The close resemblance of the cuprate phase diagram to that of other material families, such as heavy-fermions and iron-pnictides, where superconductivity occurs close to an antiferromagnetic quantum critical point (QCP) \cite{Shibauchi14} has led to speculation that the linear resistivity close to optimal doping in the cuprates may be a marker for a quantum critical transition to a hidden ordered phase \cite{Taillefer10}. The idea is that quantum fluctuations of the hidden phase provide the scattering mechanism which gives rise to both the linear resistivity and also provide the pairing mechanism for high temperature superconductivity.  Possible candidates for this order are the pseudogap or CDW, but the thermodynamic evidence for a true phase transition of any type at finite temperature is weak and there is little evidence that either of these have a quantum critical end-point \cite{Blanco-canosa14,Tabis17}.  For example, there are no observable anomalies in the specific heat \cite{Cooper14} as the material is cooled into the pseudogap or CDW regimes. On the other hand, a number of recent experiments might support the existence of a QCP close to optimal doping. First, quantum oscillations in YBa$_2$Cu$_3$O$_{6+x}$ (Y123) show that the quasiparticle mass $m^*$ increases with doping, beyond $p=0.12$,  with 1/$m^*$ extrapolating to zero at $p \sim$ 0.18 \cite{Ramshaw15}. Second, measurements suggest that in Y123,  $n_{\rm H}(0)$ undergoes a rapid increase from $p$ to 1 + $p$ over a narrow doping range 0.16 $< p <$ 0.20 \cite{Badoux16}, revealing possible critical behaviour near the doping where the pseudogap is believed to end in this material.

Whether such features are really caused by a QCP and whether this QCP is relevant to superconductivity requires further study. Recently, a high-pressure study of YBa$_2$Cu$_4$O$_8$ (Y124) \cite{Putzke16} showed that as the maximum $T_c$ is approached by pressure tuning, rather than by chemical doping, $m^*$ actually decreases. This suggests that although the mass increase in Y123 near optimal doping may be linked to quantum CDW fluctuations, these fluctuations may not be the primary cause of the high $T_c$. 

Interpreting $n_{\rm H}(0)$ as a planar hole density may have complications in some of the systems studied to date.  In Y123, the quasi-one-dimensional CuO chains layer increases the \emph{b}-axis conductivity but does not contribute to the Hall conductivity ($\sigma_{xy}$), thus increasing the measured  $n_{\rm H}(0)$ over that expected from the CuO planes alone (see SI).  In La$_{2-x}$Sr$_x$CuO$_4$ (LSCO), it is found that $n_{\rm H}(0) \simeq p$ but only for $p<0.08$ \cite{Ando04,Balakirev09}. At higher doping, $n_{\rm H}$ increases well above $(1+p)$ \cite{Ando04,Balakirev09} probably because the Fermi-surface develops electron-like curvature for doping close to $p\simeq 0.2$ where there is believed to be a Lifshitz transition \cite{Horio18}.  In Nd doped LSCO the rise in $n_{\rm H}$ with $p$ is sharper than in LSCO \cite{Collignon17} but again the interpretation may also be complicated by a change in Fermi-surface curvature \cite{Matt15}.  A detailed discussion of these issues can be found in the Supplementary information.

Here we have studied the evolution of $n_{\rm H}(0)$ in two cuprate families, Tl2201 and Bi2201, which have simple single band Fermi-surfaces and which may be overdoped to the edge of the superconducting dome or beyond without suffering Lifshitz transitions \cite{Rourke10,Ding19}.  Single crystals spanning a wide range of doping from  slightly underdoped to strongly overdoped were used in this study.  We find that $n_{\rm H}(0)$ evolves smoothly as a function of $p$ right across the overdoped regime so that $n_{\rm H}(0)$ does not reach the value $1 + p$ until close to the edge of the superconducting regime (note that our doping scales differ from some previous works as described in the SI).  The behaviour correlates well with the evolution of the linear component of $\rho_{xx}(T)$ suggesting that the two have a common origin.  Moreover there does not appear to be a simple correlation between the evolution of $n_{\rm H}(0)$ with $p$ and the closing of the pseudogap as previously conjectured \cite{Badoux16,Collignon17}.

Tl2201 has unique properties for this study.  It has a quasi-two-dimensional band-structure with a single CuO$_2$ layer giving rise to one band crossing the Fermi level.  Its maximum $T_c$ is $94$\,K and it may be sufficiently overdoped so that it becomes non-superconducting yet remains electronically sufficiently clean for quantum oscillations to be observed \cite{Vignolle08,Bangura10,Rourke10}.  For the overdoped compositions with $T_c\lesssim$26\,K, the Fermi surface geometry, scattering rate anisotropy and temperature dependence have all been accurately determined by quantum oscillation, angle-dependent magnetoresistance and angle resolved photoemission (ARPES) measurements \cite{Hussey03,Plate05, Abdel-jawad06, Abdel-jawad07,Vignolle08,Bangura10,Rourke10}. 

Bi2201 is another single-layered cuprate which has a significantly reduced maximum $T_c$ (= 34\,K) and upper critical field ($H_{c2}$), allowing superconductivity to be suppressed over a wider range of field and temperature space thereby reducing uncertainty in $n_{\rm H}(0)$. While quantum oscillations have not been observed in Bi2201, probably because it more disordered than Tl2201,  its electronic structure has nonetheless been well characterized via ARPES and scanning tunneling microscopy \cite{Kondo07,Wise08,Wise09,Kondo09,Ding19}. 

Figure 1 shows the field and temperature dependence of the Hall coefficient $R_{\rm H}$ for five representative Tl2201 samples. 
The evolution of $R_{\rm H}$ as a function of field and temperature at fixed doping can be understood, to some extent, in the overdoped regime using conventional Boltzmann transport theory.  We have calculated the expected field and temperature dependence of $R_{\rm H}$ for Tl2201 using the known Fermi surface geometry as well as the anisotropy and $T$-dependence of the scattering rate determined independently from $c$-axis magnetoresistance measurements \cite{Abdel-jawad06,French09}(see SI).  At low field, $R_{\rm H}$ is enhanced with respect to $1/ne$ (where $n$ is the carrier density determined by the Fermi surface volume) due to anisotropy in the Fermi velocity and scattering rate. At high field this anisotropy is averaged out as the electrons complete increasing fractions of their cyclotron orbits before being scattered and consequently $R_{\rm H}$ tends toward $1/ne$ (see Fig.\ S7). The low field enhancement in $R_{\rm H}$ is reduced at low temperature as scattering becomes dominated by isotropic impurity scattering and this together with the smaller scattering rate means that $R_{\rm H}$ approaches $1/ne$ at lower fields.  Higher impurity scattering increases the field scale at which $R_{\rm H}$ approaches its infinite field value, but it also considerably diminishes the enhancement of $R_{\rm H}$ over $1/ne$, so $R_{\rm H}\simeq 1/ne$ at relatively low fields.

$R_{\rm H}(H,T) $ for the most overdoped sample of Tl2201 in Fig.\ 1, with $p=0.27$, follows well the calculated behaviour. $R_{\rm H}$ at high temperature (120\,K) decreases slowly with increasing field ($\sim 10 \%$ in 60\,T), but as the temperature is lowered, this field dependence becomes weaker and for $T \leq$ 10 K, it is essentially constant once superconductivity is suppressed ($\mu_0H >$ 20\,T at $T$ = 4.2\,K). The limiting value of $R_{\rm H}$ at this doping is close to that expected for the Hall number $n_{\rm H} \sim 1 + p$.  This limiting behaviour is made clearer in the right hand panels of Fig.\ 1 where $R_{\rm H}$ is replotted against $H/\rho_{xx}^0$ which is proportional to $\omega_c\tau$ which reflects the fraction of a cyclotron orbit traversed by an electron before it scatters (here $\rho_{xx}^0$ is the extrapolated zero field resistivity).  

For the $p=0.26$ sample, at elevated temperatures the $H$ dependence of $R_{\rm H}$ becomes larger but again saturates in the high $H/T$ limit at a value consistent with $n_{\rm H} \sim 1 + p$.  For $p=0.22$, the irreversibility field has increased significantly. Nevertheless, 60\,T still appears to be sufficient to reach the limiting value of $R_{\rm H}$, though in this case, it is found to be significantly higher than that expected from $n_{\rm H} \sim 1 + p$. Indeed, for the optimally doped sample ($p=0.19$) and the slightly underdoped sample ($p=0.14$), the values of $R_{\rm H}$  at the highest field and lowest temperature correspond more closely to $n_{\rm H} \sim p$ than to $n_{\rm H} \sim 1 + p$.  For these higher $T_c$ samples, the temperature dependence of  $R_{\rm H}$ is stronger than calculated from the estimated anisotropic scattering, although $R_{\rm H}$  still decreases with increasing $H$ as expected.

For Bi2201, the field dependence of $R_{\rm H}$, above $H_{c2}$,  is much weaker than in Tl2201 (Fig.\ S3) consistent with a much higher isotropic elastic scattering rate (residual resistivities are almost one order of magnitude larger), and so we would expect the high field/low temperature values of $n_{\rm H}$ to more accurately reflect the carrier density.

The $T$-dependence of $n_{\rm H}$ at fixed high fields, for the different dopings are shown in Figure 2 for both Tl2201 and Bi2201. Extrapolating  $n_{\rm H}(T)$ to $T=0$ for each composition gives an estimate of $n_{\rm H}(0)$ whose evolution with doping is plotted in the lower panels of Figure 3. $n_{\rm H}(0)$ is found to evolve smoothly as a function of $p$ for both materials. For some doping values, multiple samples were measured and these gave consistent results, giving confidence that the error bars in $n_{\rm H}$ are accurate. For underdoped Tl2201 $n_{\rm H}(0)\simeq p$ but then increases over a broad doping range until it reaches the $n_{\rm H}(0) = 1 + p$ line approximately at $p=0.25$ ($T_c$ = 40\,K). In Bi2201, previous measurements have shown that $n_{\rm H}(0)$ follows a non-monotonic behaviour below optimal doping \cite{Balakirev03} possibly due to the presence of a CDW similar to that found in Y123 \cite{Leboeuf07}.  In the overdoped regime however, $n_{\rm H}$(0) in Bi2201, monotonically increases with increasing $p$ suggesting there are no regions of CDW order there.  By taking into account the CuO chain conductivity, we show in the SI (Fig.\ S9) that the evolution of the planar contribution to $n_{\rm H}(0)$ in Y123 may show a very similar evolution of $n_{\rm H}(0)$ with $p$ to that found here for Tl2201 and Bi2201, although further measurements of the resistance anisotropy are needed to confirm this.

One potential interpretation of our results is that the evolution in $n_{\rm H}$(0) evidences a slow closing of the pseudogap in the overdoped regime in these materials. Such behaviour has been suggested by a recent phenomenological model based on heterogeneous localization \cite{Pelc19}.  One of the clearest experimental signatures of the pseudogap is a collapse of the size of the anomaly in the electronic specific heat $\gamma$ at $T_c$ ($\Delta C(T_c)$) and a decrease in $\gamma(T)$ above $T_c$ \cite{Tallon01}.  In Tl2201, $\Delta C(T_c)$ is largest at the lowest doping measured ($p\simeq 0.20$) and $\gamma(T)$ is temperature independent above $T_c$ \cite{Wade94} which strongly suggests that $p^*<0.20$ and so the $p$ to $1+p$ transition in $n_{\rm H}(0)$ occurs in a regime where there is no pseudogap.   For Bi2201, analysis of $\rho(T)$ of our samples (see SI), suggests that $p^*<0.215$ implying that again, the transition in $n_{\rm H}(0)$ occurs at least partially in the region where there is no pseudogap.  For Bi2201 other probes suggest a larger value of $p^*$. NMR results give $p^*$ in the range 0.23-0.25 \cite{Kawasaki10}, and ARPES in the range 0.23-0.24 \cite{Kondo11, Ding19} (see SI for a discussion). Nevertheless, these estimates are still in the range where $n_{\rm H}(0) < (1+p)$.

It is possible, in principle, that a pseudogap energy scale may be below the zero field $T_c$ and this may cause the reduction in $n_{\rm H}(0)$ we see. Such a pseudogap would not be manifest in $\gamma(T)$ above or at $T_c$, but instead there should be an anomaly below $T_c$ \cite{Tallon19}.  Furthermore, this should be accompanied by an anomalous reduction in the growth of the superfluid density ($1/\lambda^2$) and $H_{c2}$ as temperature is lowered, as both are known to be strongly reduced in the pseudogap regime \cite{Tallon19}.  None of these signatures are observed experimentally in Tl2201 \cite{Wade94, Uemura93, Broun1997}. Indeed, $1/\lambda^2(T=0)$ and $\Delta C(T_c)$  are both found to be maximum at the lower doping measured ($p\simeq 0.20$), which seems to rule out this scenario in Tl2201.

In the cuprates, impurity scattering may be anisotropic \cite{Abrahams03}, arising, for example, from a region of the Fermi surface that lies close to a van Hove singularity (vHs). This could reduce $n_{\rm H}$ at low field, but not the high-field limit estimated here. Moreover, in both Tl2201 and Bi2201 all indications are that the vHs remains above the Fermi level at all doping levels studied (\cite{Ding19} and SI). With decreasing doping, the Fermi level becomes ever further removed from the vHs. Thus, anisotropic scattering is unlikely to account for the decrease in $n_{\rm H}$(0) with decreasing doping.  

The fall in $n_{\rm H}(0)$ could also be interpreted as evidence of an, as yet undetected, reconstruction of the Fermi surface which begins in the far-overdoped regime.  A reconstruction of the Fermi surface by a density wave could reduce $n_{\rm H}(0)$ \cite{Eberlein16}, however, this should also give rise to small Fermi pockets.  As quantum oscillations (QO) from the full $(1+p)$ Fermi-surface are observed in Tl2201 for $p>0.28$, the non-observation of such QO from small pockets is evidence that they do not exist. Although it is possible that the QO from these pockets could be damped if the density wave has poor coherence.

Previously, it was shown that $\rho_{xx}(T)$ in both Tl2201 and LSCO could be modelled as the sum of $T$ and $T^2$ components \cite{Hussey13}. The $T^2$ (Fermi-liquid like) component remains approximately independent of doping, whereas the anomalous linear-in-$T$ component ($\rho_{\rm lin}$) rises almost linearly with $p$ as $p$ is decreased from the edge of the superconducting dome \cite{Hussey13,Cooper09}. Recently \cite{Legros19}, $\rho_{\rm lin}$ has been associated with scattering at the so-called Planckian limit which is the maximum allowed rate at which energy can be dissipated.  Such strong scattering may derive from quasiparticle decoherence.   As shown in Figure 3, the reduction of $n_{\rm H}(0)$ appears to correlate closely with the emergence of $\rho_{\rm lin}$ in both Tl2201 and Bi2201, and so an alternative interpretation of the reduction in $n_{\rm H}$ is that it evidences a growth of quasiparticle decoherence on part of the Fermi-surface. The Hall conductivity $\sigma_{xy}$ is strongly weighted by parts of the Fermi surface with strong curvature. So if decoherence developed on the flat sections of the Fermi-surface but the quasiparticles remained coherent on the corners, this would lead to decrease in $\sigma_{xx}$, but would leave $\sigma_{xy}$ relatively unchanged, resulting in a decrease of $n_H \sim \sigma_{xx}^2/\sigma_{xy}$ as observed here.  At lower doping, once the pseudogap has developed, $n_{\rm H}(0)=p$ \cite{Ando04,Balakirev09,Badoux16,Barisic19}; a response which presumably comes solely from the remaining Fermi arcs, although the mechanism for this remains open to debate.
 
Intriguingly, for Tl2201, the region ($p>0.275$) where $n_{\rm H}(0)$ merges with the 1 + $p$ line is the only region where quantum oscillations from the full $(1+p)$ Fermi-surface have been observed \cite{Bangura10}. The quasi-classical model of angle-dependent, out-of-plane magnetoresistance, so successful in modelling the lower $T_c$ samples of Tl2201, also fails for doping less than this  ($T_c \gtrsim $ 40\,K) \cite{French09}.  Although there could be several reasons \cite{Rourke10} why quantum oscillations were not observed for higher $T_c$ samples, such as increased impurity scattering, in the light of these new results it is plausible that a loss of quasiparticle coherence around the Fermi surface is preventing quantum oscillatory phenomena from being realized.  The endpoint of the transition in $n_{\rm H}(0)$, where $n_{\rm H}(0)\simeq p$ appears to occur approximately at optimal doping (Figure 3) which is also where $\rho_{\rm lin}$ is maximum \cite{Hussey13,Cooper09}, again showing the close correlation between the two properties. If this is indeed caused by decoherence, it appears that this onsets well before  the pseudogap is evident in resistivity or specific heat.  Understanding, exactly how decoherence affects the transport properties and superconductivity could prove to be a crucial part of the high-$T_c$ cuprate puzzle.

\bibliographystyle{naturemag}
\bibliography{Tl2201_Hall}

\begin{figure*}[h]
\center
\includegraphics*[width=13cm]{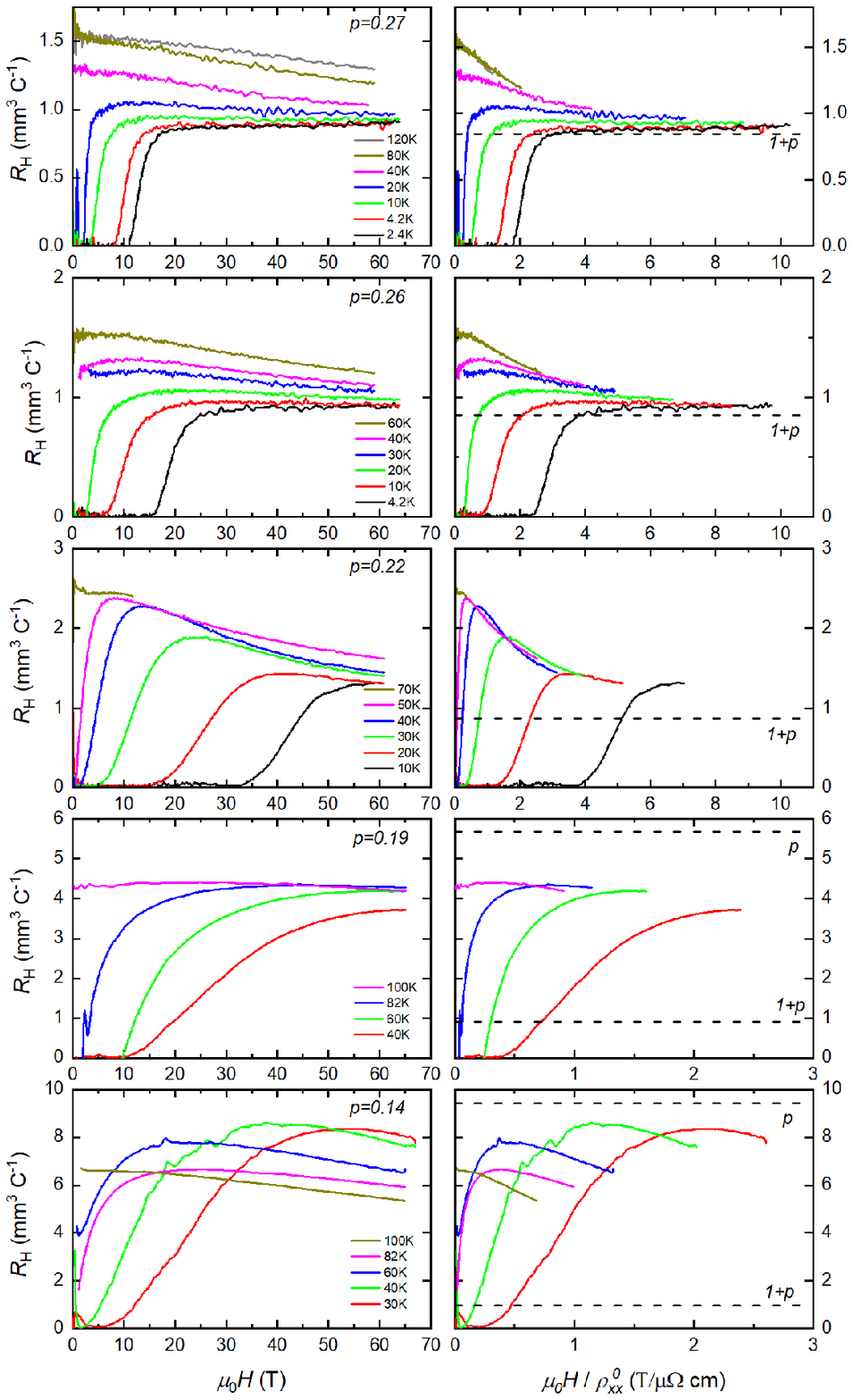}
\caption{Field dependence of the Hall coefficient $R_{\rm H}$ for samples of Tl2201 with various doping levels. The left hand panels show the raw data at different temperatures and the right panels show the same data with the field scaled by the estimated value of the resistivity $\rho(T)$ at $H = 0$. The dashed lines show the $R_{\rm H}$ values corresponding to 1/$n_{\rm H}e$ where $n_{\rm H} = 1 + p$ and $n_{\rm H} = p$.}
\label{Tl2201RHB}
\end{figure*}

\newpage

\begin{figure*}[h]
\center
\includegraphics*[width=16cm]{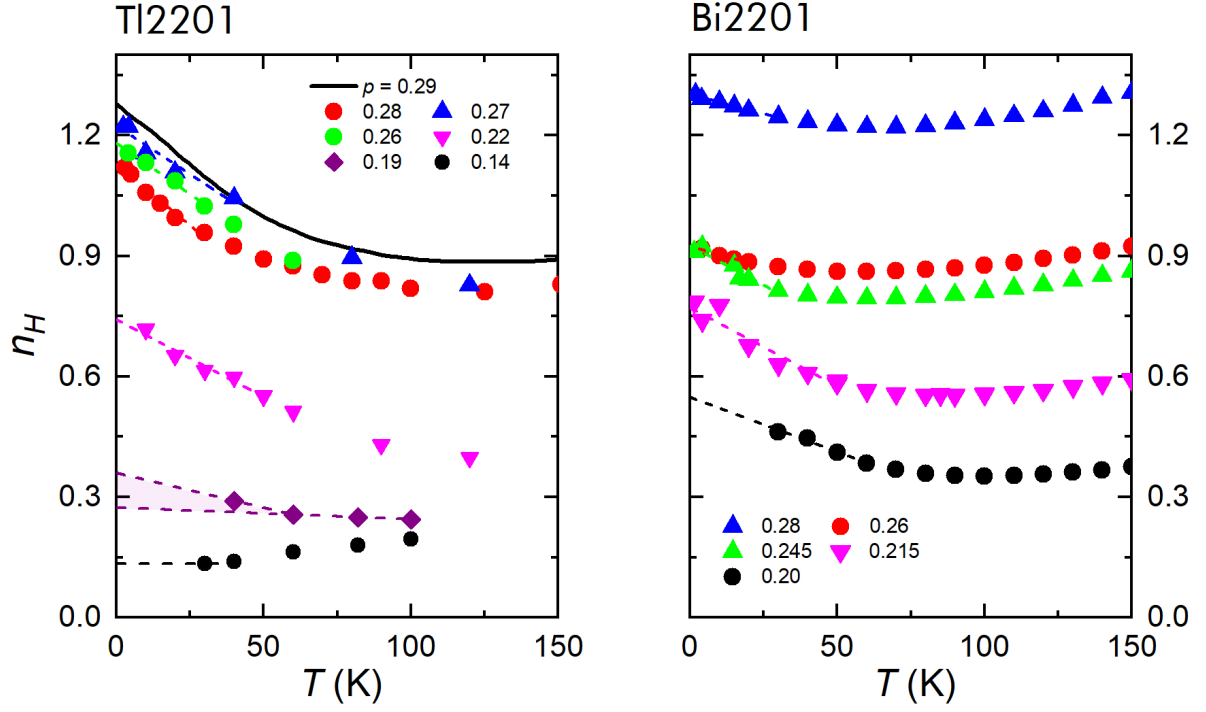}
\caption{Evolution of the Hall number $n_{\rm H}$  with temperature for samples of Tl2201 and Bi2201 with different doping values $p$ as indicated.  $n_{\rm H}$ values are taken at the highest values of field ($\mu_0H=60$ to $65$\,T) in Figure 1) expect $p=0.28$ which was measured in 14\,T.  The dashed lines show the extrapolation to $T=0$ to give $n_{\rm H}(0)$. For the sample with $p=0.19$ we show two possible extrapolations.  The solid line shows $n_{\rm H}(T)$ for a $p=0.29$ sample of Tl2201 taken from Ref.\ \cite{Mackenzie96}, with $\mu_0 H=16$\,T.}
\end{figure*}

\newpage

\begin{figure*}[h]
\center
\includegraphics*[width=16cm]{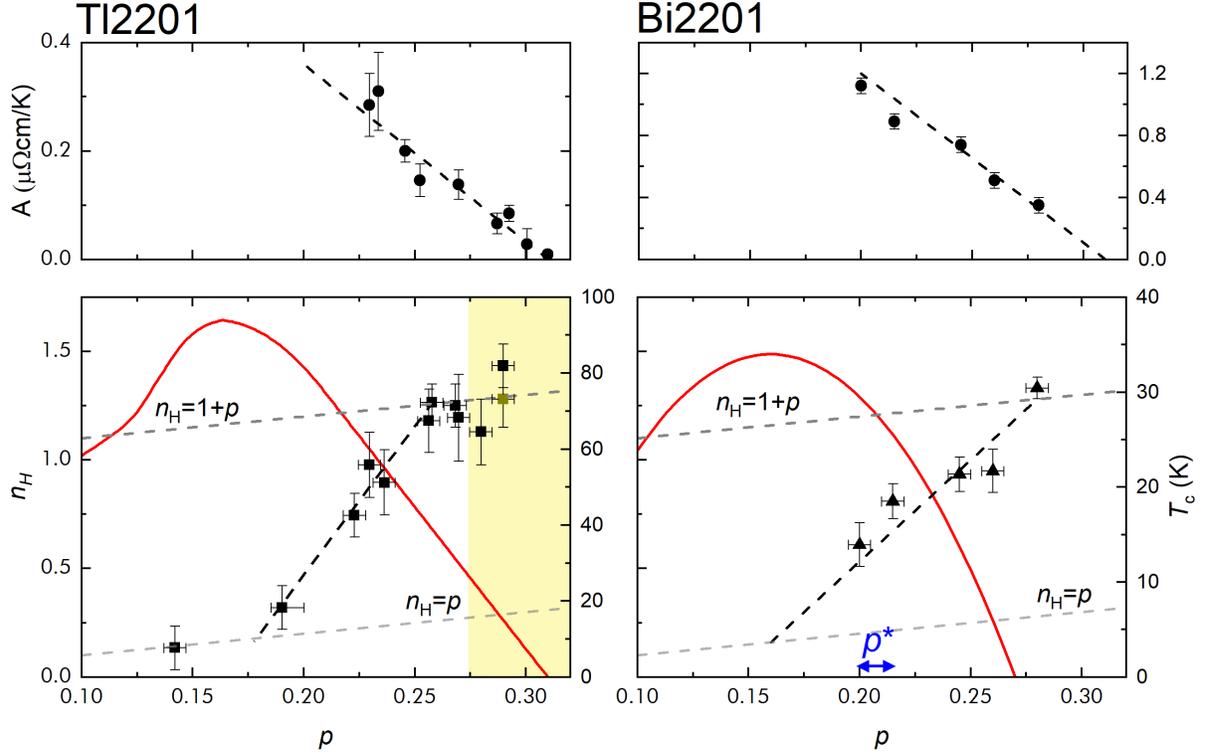}
\caption{Bottom panel: High field and low temperature values of the Hall number $n_{\rm H}(0)$ versus doped holes ($p$) for Tl2201 and Bi2201, from the extrapolations shown in Figure 2. The data point for the most overdoped sample is taken from Ref.\ \cite{Mackenzie96}. The grey dashed lines show the behaviour expected for $n_{\rm H}(0) = p$ and $n_{\rm H}(0) = 1 + p$ and the black dashed line is a guide for the eye. The error bars in $n_{\rm H}$ reflect the geometric uncertainty, whereas those for $p$ reflect the uncertainty in $T_c$.  The assumed evolution of $T_c(p)$ is shown with a solid red line for each compound (right hand scale, see SI).  The shaded area on the Tl2201 panel shows the doping range where quantum oscillations have been observed. The end of the pseudogap regime ($p^*$) as found from $\rho_{xx}(T)$ measurements in our samples is indicated ($0.20<p<0.215$).  Top Panels: Evolution of linear-in-$T$ coefficient of the zero field resistivity of Tl2201 \cite{Hussey13} and Bi2201. The $A$ error bars reflect the geometric uncertainty.}
\end{figure*}

\clearpage

\section*{Supplementary Information}
\noindent
\section{Methods}

\begin{figure} [b]
\center
\includegraphics*[width=12cm]{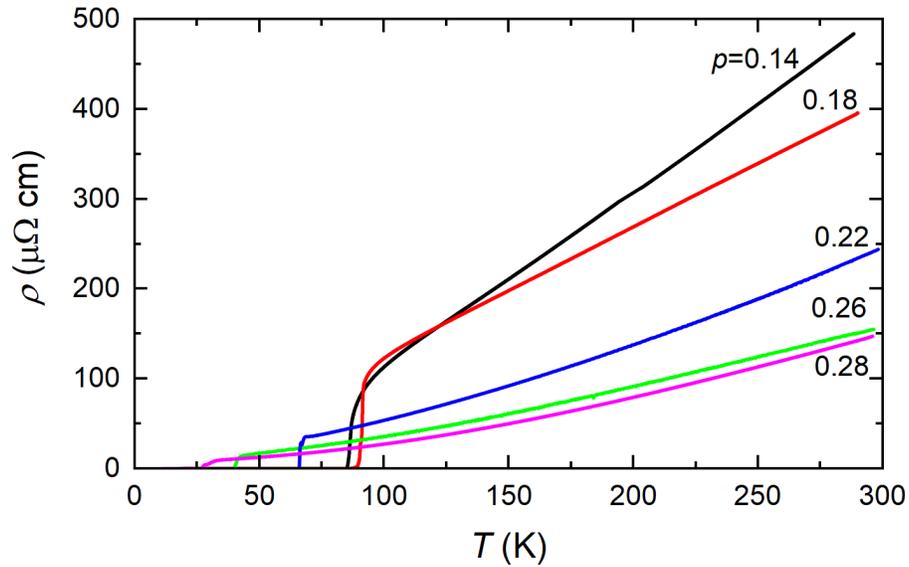}
\caption{Measured temperature dependence of the resistivity for the samples of Tl2201.}
\label{tl2201rhoT}
\end{figure}

\begin{figure} 
\center
\includegraphics*[width=10cm]{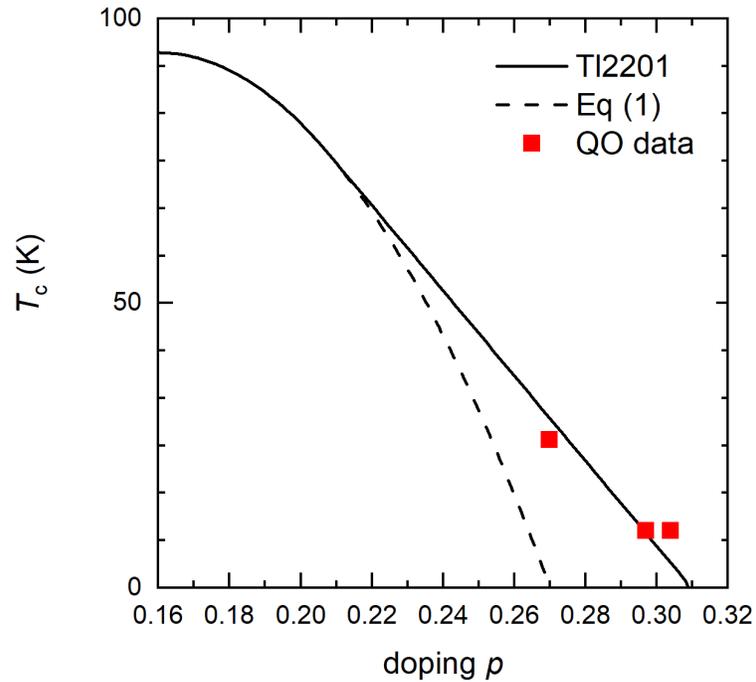}
\caption{Assumed relation between doping ($p$) and $T_c$ for Tl2201 (solid line), along with Eq.\ (\ref{EqPresland}) (dashed line) and the quantum oscillation data (square symbols)\cite{Vignolle08,Bangura10,Rourke10}.}
\label{tl2201Tcp}
\end{figure}

\noindent
\textbf{Sample preparation and characterization}\\{\textbf{Tl2201}}\\
Single crystals of Tl2201 were grown by a self flux technique using a similar technique to that described in Ref.\ \cite{tylerthesis}. We used a ZrO$_2$ crucible with a lid attached rigidly using a fused quartz holder to reduce Tl loss at high temperatures.  The samples are tetragonal with a small Cu excess and Tl deficiency.

Very thin platelet-like samples were selected, and annealed in either vacuum or an oxygen/nitrogen gas mixture for between 30 minutes and 4 hours to set the oxygen level.  Details are shown in Table S1. Following the anneal, the samples were removed from the furnace and cooled rapidly in the same atmosphere. Rapid cooling was facilitated by an aluminum heat sink around the quartz tube.   Contacts were applied by sputtering Au pads onto the top and sides of the samples using a mask. Gold wires were attached to the pads using  Dupont 4929 silver paint which was and dried at room temperature for a least one hour, prior to the oxygen-setting annealing process.

Measurements of $R_H$ were performed in pulsed field either at the LNCMI in Toulouse or the HLD in Dresden using a numerical lock-in digitization technique operating around 50\,kHz.  For each temperature point, measurements were made with the field both parallel or antiparallel to the $c$-axis. The Hall resistance $R_{xy}$ was then calculated from the odd part of the magnetoresistance and the Hall number $n_{\rm H}=V_{\rm cell}/(eR_H)$ using the cell parameters $a=3.87$\AA, $c=11.6$\AA\ which gives the cell volume for one formula unit of Tl2201. Resistance versus temperature curves  (Fig.\ \ref{tl2201rhoT}) were measured in zero field by a standard 4 probe technique using a lock-in amplifier at 72\,Hz, and were used to determine $T_c$ of the samples which we define as the temperature where the resistance falls to 1\% of the value just above $T_c$.

\setlength\tabcolsep{12pt}
\begin{table} [h]\label{tl2201details}
\begin{tabular}{|l|l|l|l|l|}
\hline
Sample & $T_c$ & doping & thickness& Anneal Conditions\\
\hline
C1 &85\,K& 0.14& 25$\mu$m& 550$^\circ$C, $10^{-5}$\,mbar\\
C2 &90\,K& 0.18& 15$\mu$m& 500$^\circ$C, $10^{-3}$\,mbar\\
C15 &65\,K& 0.22& 8$\mu$m& 550$^\circ$C, 2\% O$_2$ in N$_2$\\
678-10 &60\,K & 0.23& 18$\mu$m& 520$^\circ$C, 2\% O$_2$ in N$_2$\\
C7 &55\,K & 0.24 &13$\mu$m& 500$^\circ$C, 2\% O$_2$ in N$_2$\\
C24 &40\,K& 0.26& 16$\mu$m& 450$^\circ$C, 2\% O$_2$ in N$_2$\\
C23 &39\,K & 0.26 &30$\mu$m& 450$^\circ$C, 2\% O$_2$ in N$_2$ \\
J3 &30\,K& 0.27& 18$\mu$m& 450$^\circ$C, 2\% O$_2$ in N$_2$\\
678-19 &30\,K & 0.27& 20$\mu$m& 450$^\circ$C, 2\% O$_2$ in N$_2$\\
C20 &31\,K & 0.27 &15$\mu$m & 450$^\circ$C,  20\% O$_2$ in N$_2$ \\
C8 &22\,K & 0.28 &15$\mu$m& 450$^\circ$C, 20\% O$_2$ in N$_2$ \\
C22 &15\,K & 0.29 &50$\mu$m& 450$^\circ$C, 100\% O$_2$ \\
\hline
\end{tabular}
\caption{Physical details of the Tl2201 samples studied.}
\end{table}

\noindent
\textbf{Doping values for Tl2201}\\
The doping level of each sample was estimated from the measured $T_c$.  Conventionally, the parabolic relationship \cite{Presland91} 
\begin{equation}
1-T_c/T_{c,\rm{max}}=82.6(p-0.16)^2
\label{EqPresland}
\end{equation}
is used to relate $T_c$ to $p$, where $T_{c,\rm{max}}$ differs between compounds. This universally predicts that $T_c=0$ at $p=0.27$, which is consistent with data for La$_{2-x}$Sr$_x$CuO$_4$, where $x$ is thought to be equal to $p$.  For Tl2201 precise measurements of the Fermi surface volume are available from quantum oscillation measurements for samples with $T_c \simeq 10$\,K, and $T_c \simeq 26$\,K \cite{Vignolle08,Bangura10,Rourke10}. A linear extrapolation of this QO data suggests that $T_c$ is reduced to zero at $p\simeq 0.31$ which is a considerably higher doping than that suggested by Eq.\ (\ref{EqPresland}), possibly because $T_c$ is reduced by disorder rather than doping  in strongly overdoped La$_{2-x}$Sr$_x$CuO$_4$.  For $p\lesssim 0.21$, Eq.\ (\ref{EqPresland}) appears to be valid for Tl2201, as values of $p$ derived from thermoelectric power are consistent with other overdoped cuprates, notably Bi$_2$Sr$_2$CaCu$_2$O$_{8+x}$ (Bi2212) \cite{Obertelli92}.  Therefore for $0.16>p>0.21$ we use Eq.\ (\ref{EqPresland}) with $T_{c,\rm{max}}=94$\,K, and for higher doping a linear relation, $T_c=231.7-748p$.  As shown in Fig.\ \ref{tl2201Tcp}, this extends the $T_c$ dome consistent with the QO data, and with the linear $T_c(p)$ segment merging smoothly with the parabolic relation at lower doping without any significant discontinuity in slope.

For the Tl2201 sample with $T_c=85$\,K there is potential ambiguity as the $p$ value depends on whether this is underdoped or overdoped.  It was annealed at a higher temperature and higher vacuum level than the $T_c=90$\,K sample, both of which would be expected to produce a lower doping. The $\rho(T)$ data for this sample may also indicate underdoping as there is significant downward curvature close to $T_c$ and the absolute magnitude is higher than the $T_c=90$\,K sample at room temperature (Fig.\ \ref{tl2201rhoT}). The downward curvature could suggest the presence of a pseudogap  but could also be caused by superconducting fluctuations \cite{Tallon19}.   We therefore determined the doping of this sample ($p=0.142$) to be the same as for Y123 \cite{Liang06} with the same $T_c$ value (note $T_{c,\rm{max}}$ for Y123 is 93.6\,K which is virtually identical to that of Tl2201).  If the sample were instead overdoped, the doping would be $p=0.194$, which would move the point just to the right of the $90$\,K point in Fig.\ 3. This would not affect our conclusions significantly.

\setlength\tabcolsep{6pt}
\begin{table} [h]\label{Bi2201details}
\begin{tabular}{|l|l|l|l|l|l|l|l|}
\hline
$T_c$ & doping & thickness& Bi& Pb& Sr &La& Anneal Conditions\\
\hline
29.1\,K&0.20&6.4$\mu$m&1.20&0.90&1.30&0.55& As grown\\
25.8\,K&0.215&18$\mu$m&1.35&0.85&1.47&0.38&650$^\circ$C, 72\,h in 1 atm.\ N$_2$\\
14.5\,K&0.245&27$\mu$m&1.72&0.38&1.85&0.0&550$^\circ$C, 72\,h in 1 atm.\ N$_2$\\
6.8\,K&0.26&25$\mu$m&1.72&0.38&1.85&0.0&750$^\circ$C, 24\,h in 1 atm.\ 20\%O$_2$/N$_2$\\
$<2$\,K&0.28&44$\mu$m&1.72&0.38&1.85&0.0&400$^\circ$C, 86\,h in 2.5 atm.\ O$_2$\\
\hline
\end{tabular}
\caption{Physical details of the Bi2201 samples studied, with general formula Bi$_{2-y+z}$Pb$_y$Sr$_{2-x-z}$La$_x$CuO$_{6+\delta}$.}
\end{table}

\begin{figure}
\center
\includegraphics*[width=16cm]{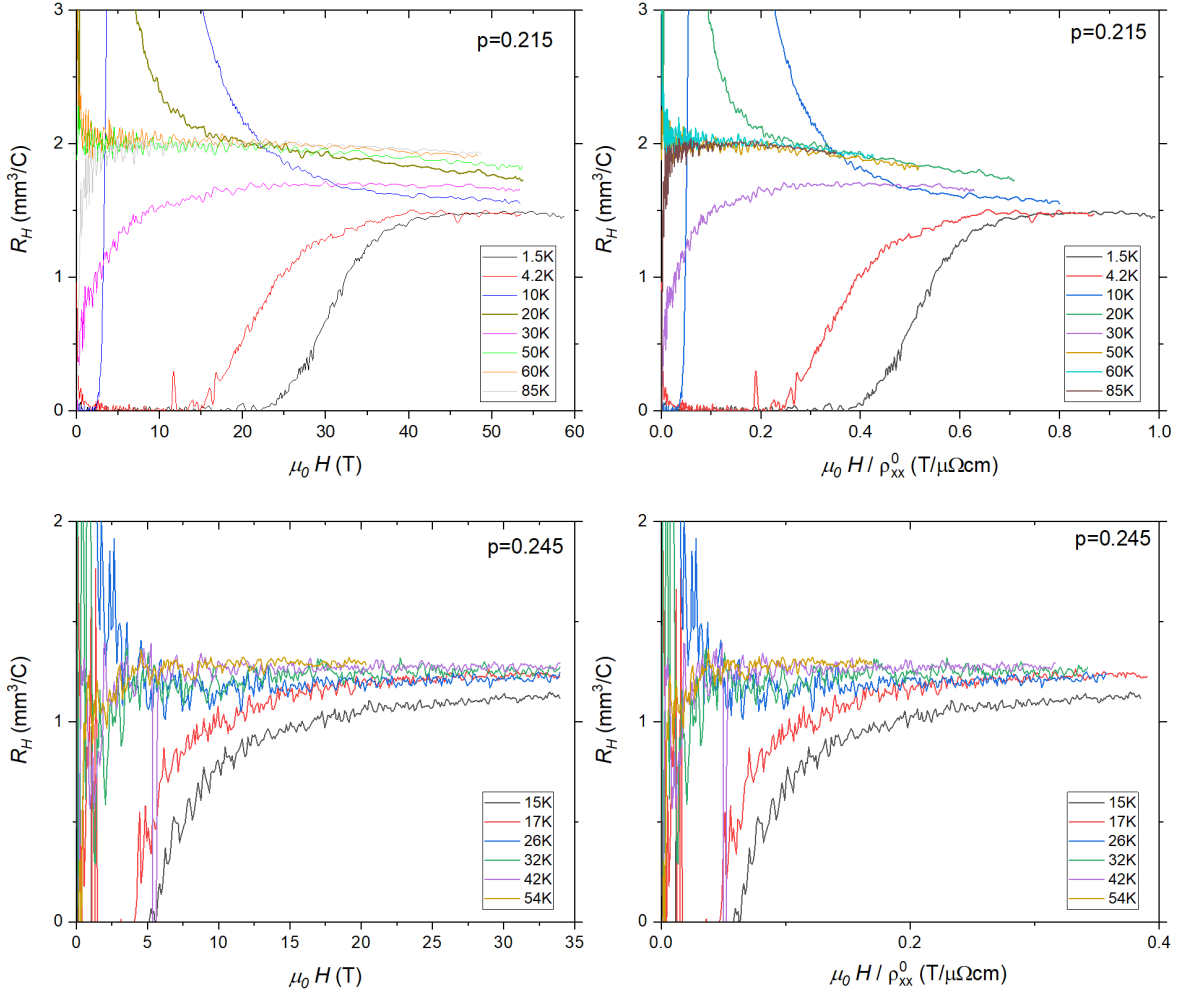}
\caption{Field dependence of $R_{\rm H}$ for the Bi2201 samples at the fixed temperatures indicated.  Two sampled with different dopings are shown as indicated in the panels. As for Figure 1 in the main text, in the right hand panels the magnetic field has been scaled by dividing by the extrapolated zero field resistance at that temperature. }
\label{SBi2201RH}
\end{figure}

\noindent
\textbf{Bi2201}\\
Single crystal samples of Bi2201 doped with both Pb and La, with general formula Bi$_{2-y+z}$Pb$_y$Sr$_{2-x-z}$La$_x$CuO$_{6+\delta}$, were grown using the floating zone technique. Samples grown in the same manner in the same laboratory have previously been studied using ARPES and STM. Table S2 lists the stoichiometry of each of our samples as well as the annealing conditions. Also listed are the $T_c$ values as determined from their resistive transitions.
Contacts were applied by gluing gold wires onto freshly cleaved surfaces using Dupont 6838 silver paint and annealing at 450$^\circ$C in flowing oxygen for 10 minutes. Zero-field resistance versus temperature curves were measured in zero field by a standard four-probe ac lock-in detection technique. Measurements of $R_{\rm H}$ were performed either in pulsed magnetic fields at the LNCMI in Toulouse or in static fields at the HFML in Nijmegen. As with the Tl2201 study, the Hall resistance $R_{xy}$ was determined by taking the anti-symmetrised signal from measurements made with the magnetic field applied both parallel and anti-parallel to the $c$-axis. Plots of the Hall coefficient versus field and temperature, similar to Fig.\ 1 of the main text, are shown in Figure\ \ref{SBi2201RH}.

\newpage
\noindent
\textbf{Doping values for Bi2201}\\
There has been some debate and uncertainty about the doping levels in Bi2201.  For our work the most important point is to determine $p^*$ in relation to the evolution of $R_{\rm H}$ and so the exact doping scale is of secondary importance.  Several previous works have used the so-called Ando scale (from Ref.\ \cite{Ando00}) which can be approximated by Eq.\ \ref{EqPresland}, but with the factor 82.6 replaced by approximately 250 (the maximum $T_c$ is still assumed to be at $p=0.16$). This causes the superconducting dome to narrow considerably. The Ando scale is based on a comparison of the room temperature $R_{\rm H}$ values for Bi2201 with those in LSCO up to $x=0.20$, and also Y123 and Tl2201 at optimal doping. Some works have extrapolated this relation to higher doping by extending the parabola \cite{He608}, or have instead extrapolated an alternative scale (also from Ref.\ \cite{Ando00}) where $p$ is linearly related to the La content in Bi2201 \cite{Zheng05}. For our work, using $R_{\rm H}$ values to set the scale would be inappropriate because the relation between $n_{\rm H}$ and $p$ is exactly what we seek to determine.  Moreover, as discussed below, LSCO undergoes a Lifshitz transition around $x =p= 0.19$ \cite{yoshida07} leading to sections of Fermi surface with negative (electron-like) curvature and as a result, a reduction in $R_{\rm H}$, even at low $T$. In Bi2201 this transition occurs at a much higher doping level close to the edge of the superconducting dome \cite{Ding19} so a difference in $R_H$ between LSCO and Bi2201 would be expected even at low temperature and it is this which has led to an effective narrowing of the $T_c(p)$ parabola in Bi2201. \textbf{In our work we have thus determined $p$ directly from Eq.\ \ref{EqPresland}, with $T_{c,max}=34$\ K.} For the sample with $T_c<2$\,K, we use the doping dependence of the linear coefficient of the resistivity ($A$) as shown in Fig.\ 3 of the main text to obtain $p = 0.28$.  

\begin{figure} [b]
\center
\includegraphics*[width=16cm]{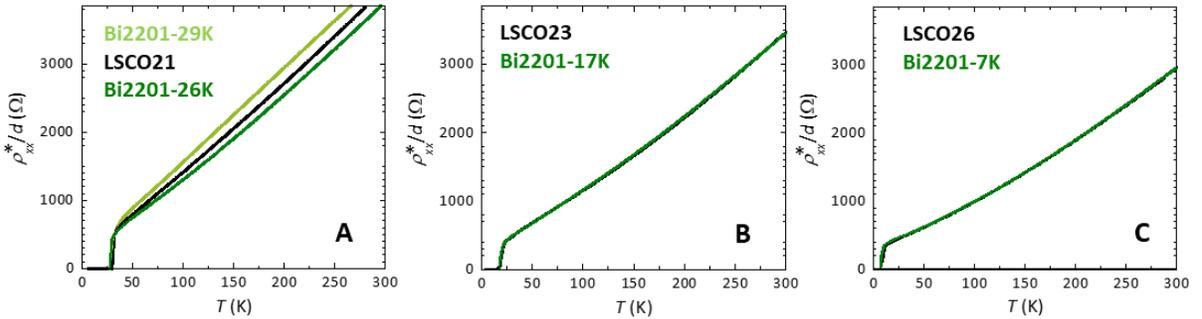}
\caption{Comparison of the zero-field resistivity in LSCO and Bi2201. {\bf A-C:} Zero-field $\rho_{xx}(T)$ curves for ({\bf A}) LSCO-21, La-Bi2201-20, La-Bi2201-21.5; ({\bf B}) LSCO-23, Bi2201-24 and ({\bf C}) LSCO-26, Bi2201-265.  For all curves, the in-plane resistivity has been divided by $d$, the $c$-axis lattice parameter (= 6.4\,\AA\ in LSCO and 12.3\,\AA\ in Bi2201). In addition, the residual resistivities and values at 300\,K have been normalised to demonstrate the similarity in the temperature dependencies and $T_c$. In panel {\bf A}, we show data for two Bi2201 samples with putative doping levels either side of $p$ = 0.20. This demonstrates that $\rho(T)$ of the LSCO ($x$ = 0.21) sample is intermediate between those of Bi2201 ($p$ = 0.20 and $p$ = 0.215).}
\label{SFrho_comp}
\end{figure}

Evidence that Eq.\ \ref{EqPresland} provides a more accurate approximation to the actual doping than the Ando scale can be found by examining the in-plane resistivity which, unlike $R_{\rm H}$ is much less sensitive to the topology of the Fermi surface and indeed, $\rho_{xx}(T)$ is found to follow a very similar evolution with doping across different cuprate families including LSCO. Fig.\ \ref{SFrho_comp} shows $\rho_{xx}(T)$ plots for LSCO and Bi2201 single crystals with comparable $T_c$ values. The resistivities show a remarkably similar $T$-dependence for all $T > T_c$. 

Further evidence can be found in STM data, where the size of the Fermi surface in Bi2201 is estimated from quasi-particle interference (QPI) \cite{He608}.   In Fig.\ (S11) of Ref.\  \cite{He608} the STM determined $p$ values are plotted against the $p$ values derived from $T_c$ either using Eq.\ (\ref{EqPresland}) or the Ando relation.  Fitting this data to a linear relation, $p_{\rm STM}=A+Bp_{T_c}$ gives $B=1.0\pm 0.1$, for Eq.\ (\ref{EqPresland}) and $B=1.9\pm 0.2$ for the Ando relation. This shows that the breadth of the $p$ scale in Eq.\ (\ref{EqPresland}) is consistent with the STM data within errors but the Ando relation is not.  Interestingly for both fits $A$ is not zero within errors [$A=0.06\pm0.02$ for the comparison with Eq.\ (\ref{EqPresland})], suggesting the STM determined dome is shifted to higher doping.  A similar discrepancy is observed in ARPES measurements \cite{Horio18} where an offset in $p$ of around 0.05 is found. In LSCO also, the ARPES determined doping is 0.10 larger than the nominal doping $x$.  The reason for this might be linked to surface effects which cause the surface to have a higher doping state, although the large escape depth of the photoelectrons in Ref.\ \cite{Horio18} may rule this out meaning the origin remains unclear.  Finally, we note that using Eq.\ \ref{EqPresland} to determine $p$ results in very similar evolution of $R_{\rm H}$(0) values in Bi2201 and Tl2201 (see Fig.\ 4 of the main manuscript). 

\begin{figure}
\center
\includegraphics*[width=14cm]{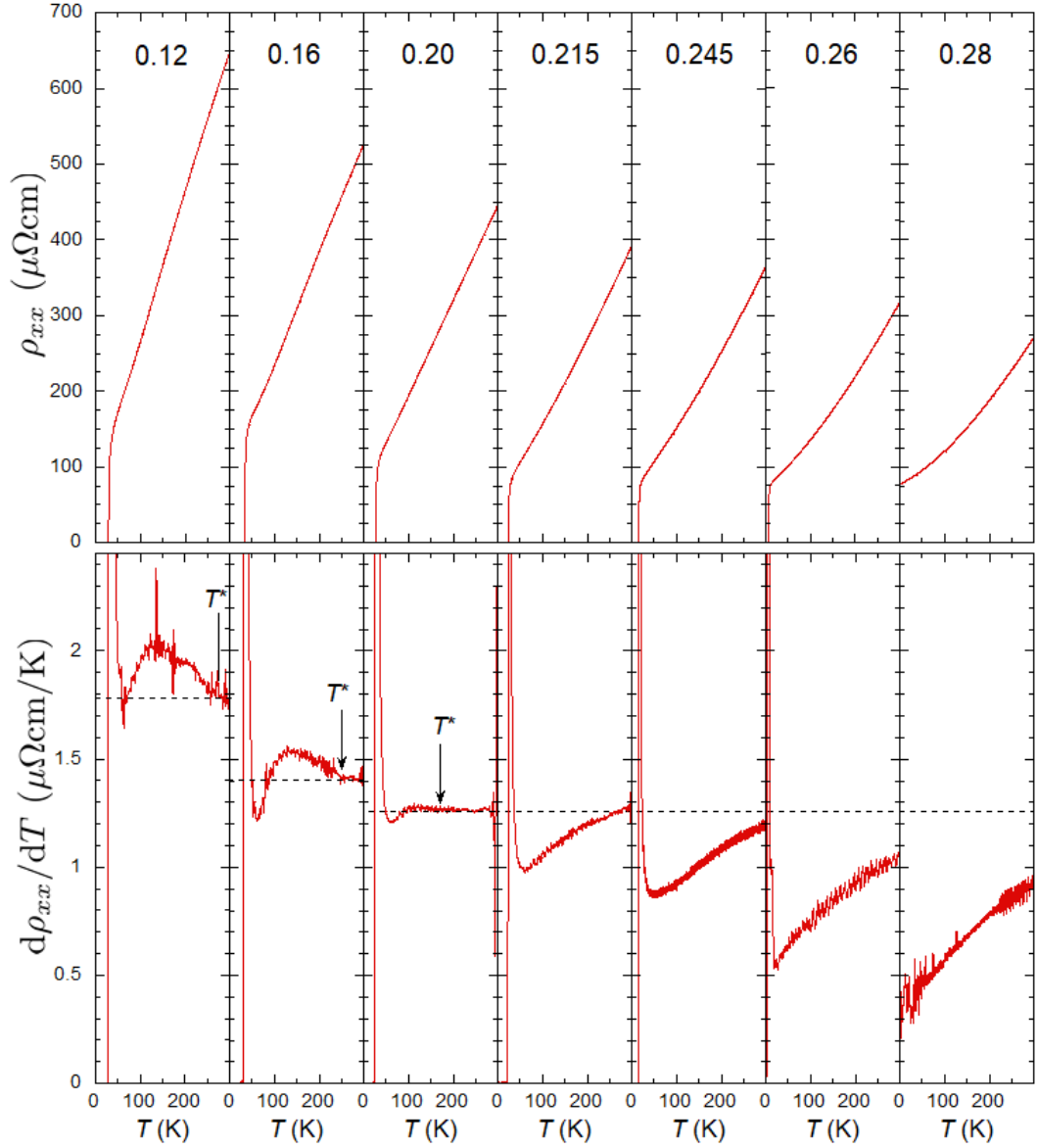}
\caption{Resistivity versus $T$ for Bi2201 samples with doping levels (determined from $T_c$ using Eq.\ \ref{EqPresland} as indicated. The bottom panels show the derivative $d\rho/dT$, with the position of the upturn due to the pseudogap indicated.  The dashed lines are guides to the eye. $T^*$ marks the temperature below which $\rho_{xx}(T)$ deviates downwards from its high-temperature $T$-linear dependence.}
\label{SBi2201RT}
\end{figure}

\bigskip
\noindent
\textbf{$T^*$ and $p^*$ values for Bi2201}\\
We now turn to consider the determination of $p^*$ in Bi2201. Given the different estimates of $p$ in Bi2201, care must be exercised when comparing values of $p^*$ between different publications.  In the present work, we have used in-plane zero-field resistivity measurements to identify the location of $p^*$ within our set of samples. The $\rho(T)$ curves for the individual Bi2201 crystals are displayed in the top panels of Fig.\ \ref{SBi2201RT}. For completeness, we show measurements over the full range of dopings that were available to us in order to highlight the location of $p^*$. In clean, underdoped cuprates, the presence of the normal state pseudogap manifests itself as a downturn in the in-plane resistivity, or equivalently an upturn in the temperature derivative d$\rho_{xx}$/d$T$, at a temperature $T^*$. As indicated in the bottom panels of Figure \ref{SBi2201RT}, upturns in d$\rho_{xx}$/d$T$ are seen for all dopings up to and including $p$ = 0.20. The upturn for $p$ = 0.20 is small but discernible.   Beyond $p$ = 0.20, however, the $\rho_{xx}(T)$ plots exhibit the characteristic super-linear shape of other overdoped cuprates without any upturns attributable to a pseudogap. The only upturns appear to be from superconducting fluctuations. This suggests that the end of the pseudogap $p^*$ is between $p=0.20$ and $p=0.215$ ($T_c=25.8$\ K) in our samples.

\begin{figure}
\center
\includegraphics*[width=8cm]{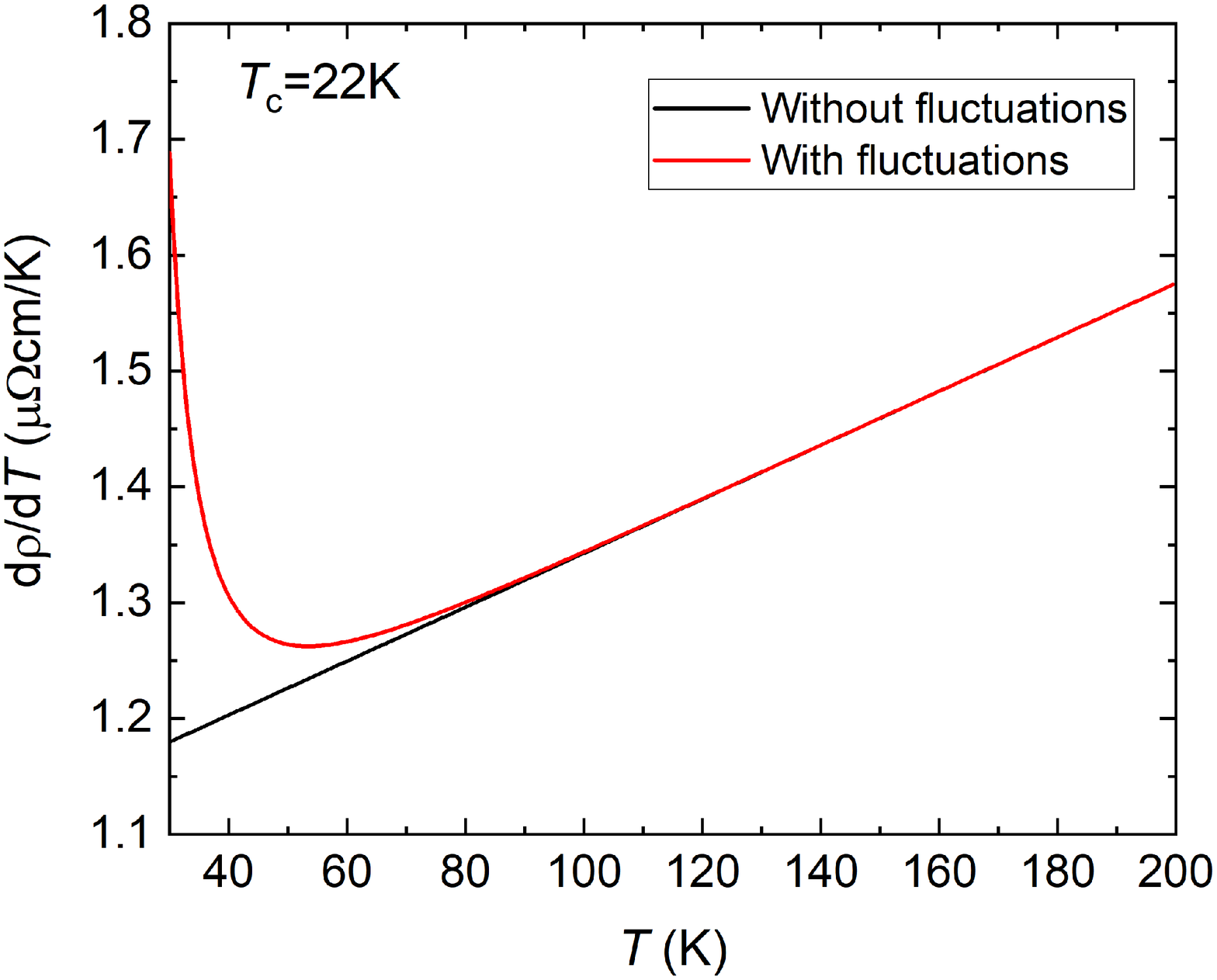}
\caption{Simulation of the effect of fluctuations on the temperature derivative of the in-plane resistivity, with parameters appropriate for an overdoped Bi2201 sample with $T_c=22$\,K.  The normal state has been approximated by $\rho(T) = \rho_0 + A T + BT^2$, with $\rho_0=80\mu\Omega$ cm, $A=1.11\mu\Omega$ cm/K and $B=2.3$\,n$\Omega$ cm/K$^2$. The fluctuations have been added using the standard Azlamazov-Larkin formula in 2D with plane spacing $d=12.3$\,\AA, and incorporating a correction to reduce the size of the fluctuations for $T\gg T_c$ as described in Ref.\ \cite{Albenque11}.
}
\label{Sfluctuation}
\end{figure}

NMR measurement of Knight shift \cite{Kawasaki10} and spin-lattice relaxation rate \cite{Zheng05} suggest that $p^*$ occurs at a slightly higher doping in Bi2201, with a sample of $p=0.23$ ($T_c$ = 20\,K) in Ref.\ \cite{Kawasaki10} showing evidence for a pseudogap ($T^*=61$\,K) where the sample with the next highest doping ($p=0.25$, $T_c=8$\,K) does not. Determinations by ARPES include Ref.\ \cite{Kondo11} where the highest doped sample ($p=0.235$, $T_c=18$\,K) was found to show a pseudogap (with $T^*=65$\,K), whereas in Ref.\ \cite{Ding19} the lowest doped sample ($T_c=17$\,K) was found to show no signs of a pseudogap.  So the exact value of $p^*$ as determined by ARPES is unclear, but still likely to be slightly higher than our determination.   With relatively sparse data points as a function of $p$, and uncertainty in the exact doping because of inhomogeneity and consequent width of superconducting transition, pinpointing the exact location of $p^*$ is difficult. Some works \cite{Kawasaki10, Kondo11} have adopted a linear extrapolation to find where $T^*(p)=0$ in Bi2201, however, in Y123, LSCO and Nd-LSCO the transition is thought to be much more sudden \cite{Cyr-choiniere18} and so a linear extrapolation will overestimate $p^*$.  The highest doped samples in both the ARPES and NMR studies which show a pseudogap are not significantly higher than our value of $p^*$.  The NMR ($T_c=8$\,K \cite{Kawasaki10}) and ARPES studied samples ($T_c=17$\,K \cite{Ding19}) which do \textbf{not} show a pseudogap have dopings well before the end of the superconducting dome where we find $n_{\rm H}$ is still reduced below $1+p$ (Fig.\ 4 main text) and so qualitatively our conclusions are not challenged.   We recall that in Tl2201 (main text), there is very robust evidence from the large size of the specfic heat anomaly and high superfluid density, that there is no pseudogap for $p>0.20$.

An additional problem is the influence of superconducting fluctuations which, like the pseudogap, also cause an downturn in $\rho(T)$ and a suppression of the density of states close to the Fermi-level. This downturn also suppresses the ARPES and NMR intensities. This problem has been discussed by Tallon \textit{et al.} \cite{Tallon19} and demonstrated experimentally in Ref.\ \cite{Rourke11} for $\rho(T)$.  In Fig.\ \ref{Sfluctuation}, we simulate the effect of fluctuations on d$\rho/$d$T$ using parameters appropriate to a $T_c=22$\,K overdoped Bi2201 sample \cite{Kondo11}.  It can be seen that the upturn in d$\rho/$d$T$, which could be erroneously attributed to a pseudogap,  occurs at $T\simeq 70$\,K which is around 3 times $T_c$. As the Azlamazov-Larkin fluctuation contribution to the conductivity depends only on the interlayer-spacing, its relative effect is much larger in materials with high residual resistivities (such as Bi2201) compared to samples with low residual resistivities such as Tl2201.  

\newpage
\noindent
\section{Simulation of field and temperature dependence of $R_{\rm H}$ within Boltzmann transport theory}

\begin{figure} [b]
\center
\includegraphics*[width=16cm]{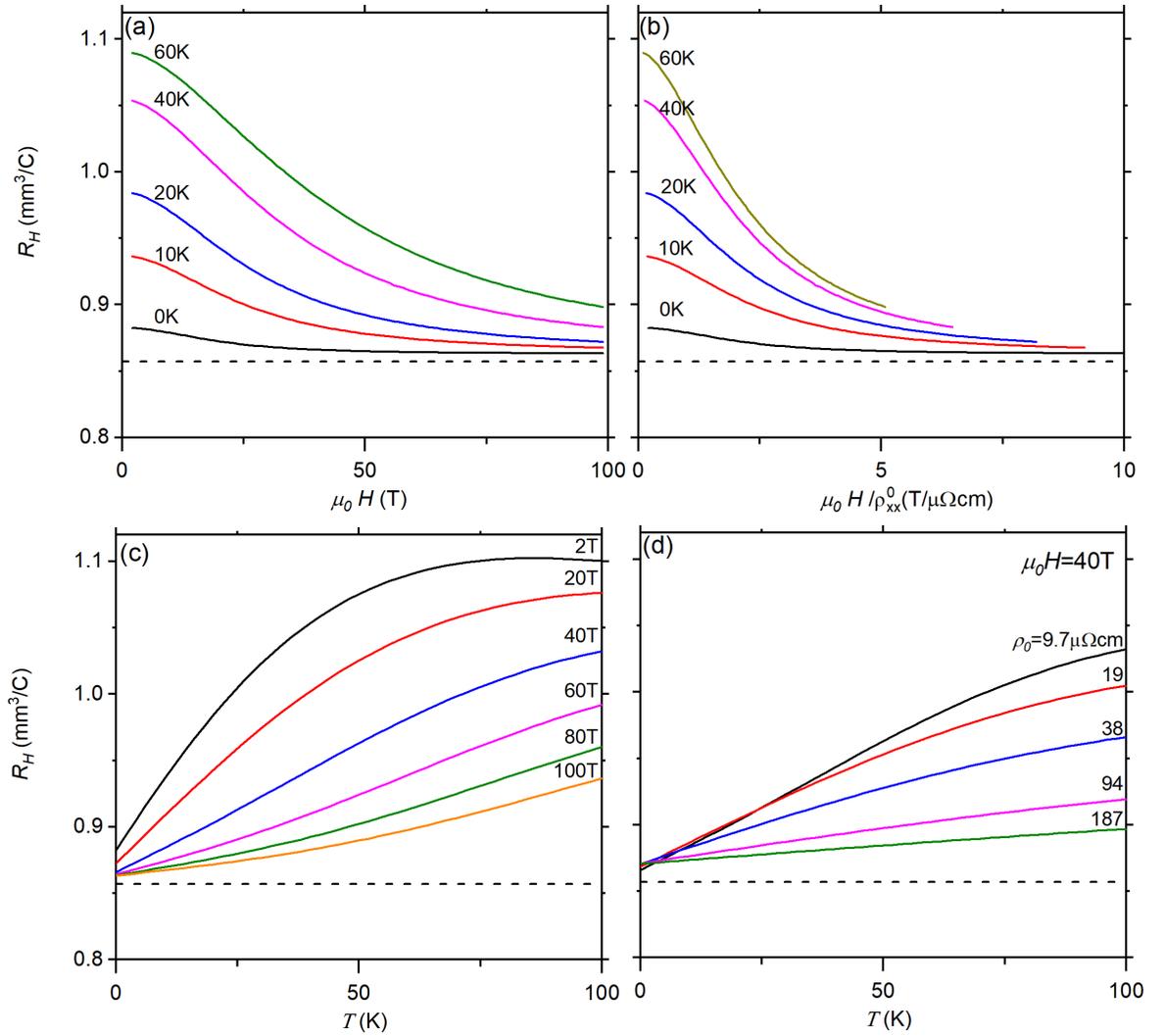}
\caption{Simulation of $R_{\rm H}$ for Tl2201, with $T_c=20$\,K. (a) Field dependence of $R_{\rm H}$ at fixed temperatures as indicated. (b) Akin to Kohler's plot $R_{\rm H}(H)/\rho_{xx}^0$. (c)  Temperature dependence of $R_{\rm H}$ at the fixed fields indicated. (d) Effect of increased impurity scattering on the temperature dependence of $R_{\rm H}$ at a fixed field of $\mu_0H=40$\,T.  The dashed line in all panels indicates the value of $R_{\rm H}$ calculated from the expression $R_{\rm H}=1/ne$, where $n= \pi k_{00}^2/(2\pi c)$. }
\label{SFigSim1}
\end{figure}

To understand the evolution of $R_{\rm H}$ with field and temperature, we calculated the expected behaviour using conventional Boltzmann transport theory and the Fermi surface parameters of Tl2201 which have previously been determined from angle dependent $c$-axis magnetoresistance (ADMR) measurements. In general, the conductivity tensor can be calculated from the Shockley-Chambers tube-integral formula (SCTIF), generalized for the case where both the Fermi velocity $v_f$ and the scattering rate vary around the Fermi surface \cite{Abdel-jawad06}.  This is valid for arbitrary field values (i.e., not only in the low or high field limits). For a two-dimensional Fermi surface, the SCTIF is given by
\begin{equation}
\sigma_{ij} = \frac{e^3 B}{2\pi^2\hbar^2 c}\int_0^{2\pi} d\phi \int_0^\infty d\phi^\prime \frac{v_i(\phi)v_j(\phi-\phi^\prime)}{\omega_c(\phi)\omega_c(\phi-\phi^\prime)}e^{G(\phi-\phi^\prime)-G(\phi)}
\end{equation}
where $G(\phi)=\int d\phi/[\omega_c(\phi)\tau(\phi)]$, $c$ is the $c$-axis lattice constant and ${i,j} \in (x,y,z)$ and $\phi$ is the azimuthal angle.  For Tl2201, the ADMR data was fitted by assuming $k_F(\phi)=k_{00}+k_{40}\cos(4\phi)$, $\tau(\phi)=\tau^0/(1+\gamma\cos 4\phi)$; i.e., the first two terms in a cylindrical harmonic expansion of $k_F(\phi)$ and the four-fold angle dependence to the scattering rate consistent with the crystal symmetry. In the fits, it was not possible to discriminate between the angle dependence of  $\omega_c$ and $\tau$ so $\omega_c(\phi)$ was set to be independent of $\phi$, and varies linearly with $H$ ($\omega_c = \omega_c^0 H$) \cite{Abdel-jawad06}. So the relevant parameters are $k_{00}$, $k_{40}$, $\omega_c^0\tau^0 (T)$ and $\gamma(T)$.  Within this approximation
\begin{equation}
\sigma_{ij} =  \frac{e^3 B}{2\pi^2\hbar^2c \omega_c^2}\int_0^{\pi} d\phi \int_0^\infty d\phi^\prime v_j(\phi)v_j(\phi-\phi^\prime)e^{G(\phi-\phi^\prime)-G(\phi)}
\end{equation}
where $v_x=k_f(\phi) \omega_c \cos(\phi-\zeta) \hbar/(eB)$, $v_y=k_f(\phi) \omega_c \sin(\phi-\zeta) \hbar/(eB)$ and $\zeta$ is the angle between $k_f$ and $v_f$; $\tan (\zeta) = d\ln k_f(\phi)/d\phi$. Both $\sigma_{xx}$ and $\sigma_{xy}$ are calculated for each temperature point and then $R_{\rm H}=-\sigma_{xy}/(\sigma_{xx}^2+\sigma_{xy}^2)B$.  $\omega_c^0\tau^0(T)$ and $\gamma(T)$ are related to the anisotropic and isotropic components of the scattering rate which were found to have linear and quadratic temperature dependencies respectively; $(1-\gamma)/(\omega_c^0\tau^0) = A_{iso}+B_{iso}T^2$ and $2\gamma/\omega_c^0\tau^0=A_{ani}+B_{ani}T$.

\setlength\tabcolsep{12pt}
\begin{table} [h] \label{Tl2201RHsim_params}
\begin{tabular}{|r|l|}
\hline
Parameter&Value\\
\hline
$k_{00}$&0.728 \AA$^{-1}$\\
$k_{40}/k_{00}$&$-3.3\times 10^{-2}$\\
$A_{iso}$ & $2.43$\\
$B_{iso}$ & $3.13\times 10^{-4}$\\
$A_{ani}$&$0.135$\\
$B_{ani}$&$6.11\times 10^{-2}$\\
\hline
\end{tabular}
\caption{Parameters derived from ADMR fits used for the magneto-transport calculations (sample tl200c1 from Ref.\ \cite{Mattthesis}) valid for $T\leq 70$\,K and at $\mu_0H$=45\,T.}
\end{table}

Results of the calculations using the parameters derived from ADMR fits to a $T_c=20$\,K sample of Tl2201 \cite{Mattthesis} (parameters given in Table \ref{Tl2201RHsim_params}), are shown in Figure \ref{SFigSim1}.  In panel (a) of Figure \ref{SFigSim1} the field dependence of $R_{\rm H}$ at different temperatures is seen to closely resemble the experimental behaviour seen in Figure 1 of the main text. There is a monotonic decrease of $R_{\rm H}$ as the field is increased with $R_{\rm H}$ tending towards $1/ne$ at the highest fields.  This is emphasized in panel (b) of the same figure.   In simple metals, all components of the magnetoresistance tensor should fall onto a common curve when plotted as a function of $\omega_c\tau$, and which is proportional to the $H/\rho_{xx}^0$, where $\rho_{xx}^0$ is the longitudinal resistance in zero field.  In Tl2201, this behaviour, which is known as Kohler's rule, is not expected because the \emph{anisotropy} of the scattering rate is temperature dependent.  The $R_{\rm H}(H)$ values do not collapse onto a common curve but can be seen to approach a common limiting value at high field.  At the lowest temperature, there remains some small field dependence of $R_{\rm H}$ because of the non-circular shape of the Fermi surface.  $R_{\rm H}$ does not exactly equal $1/ne$ for either an isotropic scattering rate or an isotropic mean-free-path in the low field limit, however, the deviations for the Fermi surface parameters of Tl2201 are small.

In panel (c) of Figure \ref{SFigSim1}, we show how the temperature dependence of $R_{\rm H}$ evolves with field.  At higher field the temperature dependence of $R_{\rm H}$ is much weaker, and it is easier to extrapolate to the zero temperature limit where $R_{\rm H}$ is close to $1/ne$.  Finally, in panel (d) we show the effect of increased impurity scattering which is relevant for the interpretation of the $R_{\rm H}$ data of Bi2201.  Here we have increased $A_{iso}$ which has the effect of increasing the residual resistance value $\rho_0$.  The figure shows that at a fixed field of 40\,T the effect of the increased impurity scattering is to strongly suppress the temperature dependence of $R_{\rm H}$.  Hence, although the values of $\omega_c\tau$ will be lower for less pure samples, the value of $R_{\rm H}$ is still close to $1/ne$.

\newpage
\section{Lifshitz transition}
\begin{figure}
\center
\includegraphics*[width=8cm]{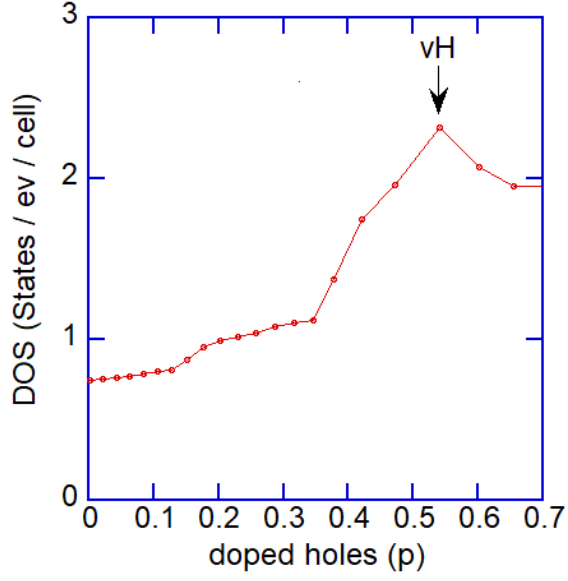}
\caption{DFT calculation of density of states for Tl2201, plotted versus doping $p$. The van Hove point where the Fermi surface undergoes a Lifshitz transition is marked vH.}
\label{tl2201vh}
\end{figure}

Density functional theory (DFT) has been found to provide an excellent description of the electronic structure of Tl2201 close to the Fermi level \cite{Rourke10}.  In Fig.\ \ref{tl2201vh} we show the density of states of Tl2201, calculated using DFT as described in Ref. \cite{Rourke10}, plotted against doping within a rigid band model.  At low doping, the Fermi surface is closed around the $M$ point in the Brillouin zone. For $p\gtrsim 0.5$ however, the Fermi surface becomes closed around the $\Gamma$ point. This Lifshitz transition results in a peak in the density of states and is labelled vH in the figure.  A similar Lifshitz transition occurs in all cuprates, but unlike in LSCO and Nd-LSCO, in Tl2201 it occurs for $p$ much larger than the range for which superconductivity is observed.  This is consistent with the fact that $R_{\rm H}$ remains positive throughout the doping range we have measured in our study.

\newpage
\noindent
\section{Hall number in other cuprates}
Here we discuss previously reported data for the Hall number versus doping in two other important cuprate superconductors;YBa$_2$Cu$_3$O$_{6+x}$ (Y123) and LSCO (with and without Nd doping) and its relation to our results for Tl2201 and Bi2201.

\begin{figure}[b]
\center
\includegraphics*[width=10cm]{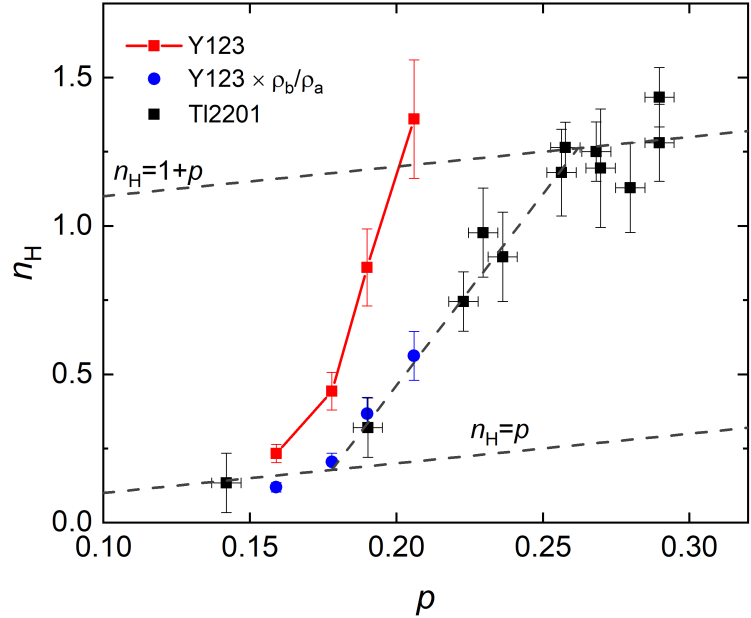}
\caption{Hall number versus doping for YBa$_2$Cu$_3$O$_{6+x}$ at $T=50$\,K along with the same data scaled by the resistivity anisotropy $\rho_b/\rho_a$ measured at $T=100$\,K \cite{Ando02}. Here we have converted the oxygen content values given in Ref.\ \cite{Ando02} into the equivalent $p$ values using the data in Ref.\ \cite{Liang06}. The most highly doped sample ($p=0.206$) is Ca-doped Y123 for which there is no corresponding anisotropy data in Ref.\ \cite{Ando02}, so we used the value for the highest doping in Ref.\ \cite{Ando02} $(p=0.194)$. The data for Tl2201 from Fig.3 (main text) are shown for comparison. The dashed lines show the relations $n_{\rm H}=p$ and $n_{\rm H}=1+p$ along with a guide to the eye through the Tl2201 data (as in Fig. 3 (main text)).}
\label{SYBCOTl2201nH}
\end{figure}

In Figure \ref{SYBCOTl2201nH} we show the reported data for $n_{\rm H}$ of Y123 \cite{Badoux16} alongside the present data for Tl2201. For Y123, $n_{\rm H}$ at 50\,K is shown because the field was not sufficiently high to suppress superconductivity to lower temperatures for all dopings, and $n_{\rm H}$ appears to be weakly $T$-dependent below this.  Unlike the smooth evolution in Tl2201, the Y123 data seem to show a very sharp rise around $p=0.19$, suggesting a phase transition which corresponds to the value of $p$ where the pseudogap is thought to close \cite{Badoux16}. However, as remarked in the main text, Y123 has a much more complicated electronic structure than Tl2201, with two unequally doped quasi-two-dimensional CuO$_2$ plane sheets of Fermi surface together a quasi-one-dimensional CuO chain sheet \cite{Pickett90}. According to DFT calculations the chain and the outer-plane Fermi surface sheets hybridize. The chain sheet is believed to be highly conducting when the composition is close to YBa$_2$Cu$_3$O$_{7}$ and responsible for the large anisotropy observed in the normal state resistivity \cite{Ando02}. The anisotropy $\rho_a/\rho_b$ is strongly dependent on the oxygen content, at $T=300\,$K varying between $2.7$ for $x=1.0$ to $1.3$ for $x=0.6$.

The effect of the chains on the Hall resistivity can be estimated by approximating them as purely one-dimensional conductors in parallel with the two-dimensional planes.  In this case, the chains will contribute to the conductivity in the $b$-direction only and will have zero Hall-conductivity $\sigma_{xy}$.  By adding the plane and the chain conductivity tensors, it is found that the Hall number coming from the planes alone $n_{\rm H}^p$ is simply related to the measured Hall number $n_{\rm H}$ by
\[
n_{\rm H}^p = \frac{\rho_b}{\rho_a}n_{\rm H}.
\]
Unfortunately, as far as we are aware, there is no reported data for the normal state resistivity anisotropy ($\rho_b/\rho_a$) at low temperature and high fields ( $T=50$\,K, $\mu_0H>70$\,T) where the Hall effect data were collected.  Using the values at the lowest temperature for which data is available ($T=100$\,K) and zero field \cite{Ando02}, results in values for $n_{\rm H}^p$ which are quite consistent with our data for Tl2201 (Figure \ref{SYBCOTl2201nH}).   $\rho_b/\rho_a$ is however, found \cite{Ando02} to be quite temperature dependent, decreasing with decreasing temperature, and for the most overdoped sample in Ref.\ \cite{Badoux16} $n_{\rm H}$ was found to increase slightly with decreasing temperature.  Both of these effects would increase $n_{\rm H}^p$ for Y123 above the Tl2201 data.  So despite the apparent consistency of the scaled $n_{\rm H}^p$ data for Y123 with that for Tl2201 shown in Figure \ref{SYBCOTl2201nH}, further experiments are required to evaluate the true behaviour for $n_{\rm H}$ for Y123.

\begin{figure}
\center
\includegraphics*[width=12cm]{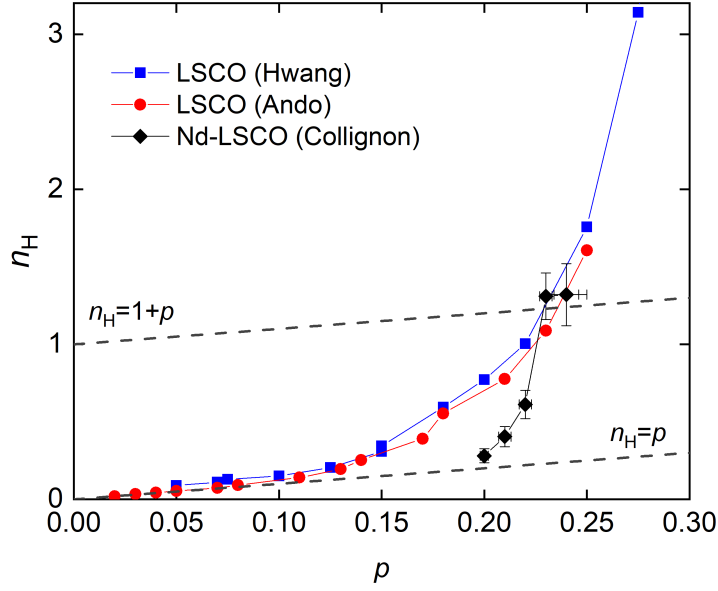}
\caption{Hall number versus doping for LSCO from Hwang et al. \cite{Hwang94} and Ando et al.\cite{Ando04}, and for Nd-LSCO \cite{Collignon17}.  For LSCO the data is taken at $T=50$\,K to avoid the effects of superconductivity or localization, whereas for Nd-LSCO $n_{\rm H}$ is extrapolated to $T=0$.}
\label{SLSCOnH}
\end{figure}

In Figure \ref{SLSCOnH} we show the reported data for $n_{\rm H}$ of LSCO \cite{Hwang94,Ando04}. There is good agreement between the two data sets shown and also a third study \cite{Balakirev09} which extends measurements to higher field and lower temperature (but has a more limited range of doping).   For LSCO at low doping ($x<0.08$) $n_{\rm H}$ follows closely the $n_{\rm H}=p$ line but for higher dopings $n_{\rm H}$ deviates strongly upwards monotonically with no discernable feature as it crosses the $n_{\rm H}=1+p$ line, reaching $n_{\rm H}=15$ at $p=0.34$.  The high field, $T=0$ limit data from Ref.\ \cite{Balakirev09} is qualitatively similar but shows an even stronger increase with doping beyond $x=0.08$.  For large $p$ in LSCO, it is likely this behaviour is driven by an approach to a Lifshitz transition, where the hole like Fermi-surface (closed around the M point) at lower doping switches to an electron-like (closed around the $\Gamma$ point) at high doping. Although all cuprates are expected to have such a transition, it is unclear in LSCO exactly at which doping it occurs.  Recent ARPES measurements \cite{Horio18} suggest it occurs around $x=0.2$.   In principle, such a Lifshitz transition is first-order as a function of $p$ at $T=0$ and should produce a switch in the sign of $R_{\rm H}$ without it passing through zero.  However, experiments suggest that the change is more gradual, possibly because of disorder and the presence of anistropic scattering \cite{Narduzzo08}, resulting in a divergence in $n_{\rm H}$.   The increase in $n_{\rm H}$ for $p$ slightly greater than $0.08$ is unlikely to be a precursor to the Lifshitz transition but is more likely to result from the change in Fermi-surface associated with the formation of a charge-density-wave (CDW) which is known to form over a range of doping centered on $p=0.12$ \cite{Croft14}. Such a reconstruction should produce small electron-like and hole-like Fermi surface pockets. In Y123 this results in $R_{\rm H}$ becoming negative for $0.08<p<0.16$ \cite{Leboeuf11,Badoux16} as $T\rightarrow 0$, presumably due the mean-free-path on the electron-like pockets becoming larger than on the hole pockets (the latter are likely to be more affected by the pseudogap).  In LSCO, the higher degree of disorder relative to Y123 probably results in the mean-free-path being less enhanced on the electron-pockets and so $R_{\rm H}$ is enhanced but does not become negative. In Ref.\ \cite{Badoux16X} $R_{\rm H}$ is found to become negative for dopings close to $x=0.12$ which might reflect the higher quality of those samples (longer electron-pocket mean-free-path) compared to previous reports.

In Figure \ref{SLSCOnH} we also show data for Nd-LSCO \cite{Collignon17} which contrasts sharply with that for LSCO. Doping LSCO with Nd, reduces $T_c$ significantly and induces static spin-charge stripe order up to $x=0.20$ in zero field \cite{Tranquada97}. Specific heat measurements suggest that the Lifshitz transition occurs between $p=0.23$ and $p=0.25$ \cite{Michon19}. High magnetic field will likely enhance this static spin-charge stripe order.  There is a strong increase in $n_{\rm H}$ as doping is increased above $p=0.20$, with $n_{\rm H}$ becoming approximately equal to $1+p$ at $p=0.23$, i.e., the Lifshitz doping transition point.  A crucial question is whether it indeed saturates for $p>0.23$  or would continue to rise as in LSCO if data for higher doping were available. Given the expected change in sign of $n_{\rm H}$ for higher doping, it seems likely that it should increase strongly as in LSCO.   Note that the data point marked at $p=0.24$ has a larger error in doping than the rest.  The authors found that the doping inferred from $T_c$ ($p=0.236\pm0.002$) was substantially lower than that from the Sr content $p=0.25$, so for such high dopings values it seems that the simple relation $p=x$ breaks down.

In summary, the data for Y123 when scaled to take into account the likely chain conduction may be consistent with the present data for Tl2201 and La-Bi2201, although there is uncertainty in the scaling factor partly due to lack of relevant experiment data.  For LSCO, the evolution of $n_{\rm H}$ with $p$ seems to be dominated by a CDW phase for doping $0.16>p>0.08$ and then the approach towards a Lifshitz transition as $p\rightarrow 0.2$ making the relation of $n_{\rm H}$ to the carrier concentration is very complicated. For Nd-LSCO, data for $p>0.24$ is required to establish if $n_{\rm H}$ does saturate at $n_{\rm H}=1+p$ or instead changes sign as expected from the apparent Lifshitz transition between $p=0.23$ and $p=0.25$.

\end{document}